\documentstyle[12pt]{article}
\newfont{\goth}{eufm10 scaled\magstep1}
\newfont{\bbbold}{msbm10 scaled\magstep1}
\newcommand{\com}{\mbox{\bbbold C}}

\newcommand{\mink}{\mbox{\bbbold M}}
\newcommand{\proj}{\mbox{\bbbold P}}
\newcommand{\gog}{\mbox{\goth g}}
\newcommand{\gp}{\mbox{\goth p}}
\newcommand{\gb}{\mbox{\goth b}}
\newcommand{\gh}{\mbox{\goth h}}
\newcommand{\gl}{\mbox{\goth l}}
\newcommand{\x}{\times}
\newcommand{\df}[3]
{\begin{picture}(390,145)(0,-15)
\put(185,80){\vector(-1,-1){50}}
\put(205,80){\vector(1,-1){50}}
\put(195,90){\makebox(0,0)[b]{#1}}
\put(125,20){\makebox(0,0)[t]{#2}}
\put(265,20){\makebox(0,0)[t]{#3}}
\put(195,16){\makebox(0,0)[t]{$\Longleftrightarrow$}}
\put(150,60){\makebox(0,0){$\pi_1$}}
\put(240,60){\makebox(0,0){$\pi_2$}}
\end{picture}}
\newcommand{\dfl}[3]
{\begin{picture}(390,145)(0,-15)
\put(185,80){\vector(-1,-1){50}}
\put(205,80){\vector(1,-1){50}}
\put(195,90){\makebox(0,0)[b]{#1}}
\put(105,20){\makebox(0,0)[t]{#2}}
\put(265,20){\makebox(0,0)[t]{#3}}
\put(195,16){\makebox(0,0)[t]{$\Longleftrightarrow$}}
\put(150,60){\makebox(0,0){$\pi_1$}}
\put(240,60){\makebox(0,0){$\pi_2$}}
\end{picture}}
\newcommand{\sif}[2]
{\begin{picture}(390,145)(0,-15)
\put(195,80){\vector(0,-1){50}}
\put(195,90){\makebox(0,0)[b]{#1}}
\put(195,20){\makebox(0,0)[t]{#2}}
\put(205,60){\makebox(0,0){$\pi$}}
\end{picture}}
\newcommand{\f}[1]{{\bf #1}}
\newcommand{\r}[1]{{\rm #1}}

\newcommand{\bttm}{\left(\begin{array}{cc}}
\newcommand{\ettm}{\end{array}\right)}
\newcommand{\mb}[1]{\left(\begin{array}{#1}}
\newcommand{\me}{\end{array}\right)}
\newcommand{\beq}{\begin{equation}}
\newcommand{\eeq}{\end{equation}}
\newcommand{\beqa}{\begin{eqnarray}}
\newcommand{\eeqa}{\end{eqnarray}}
\newcommand{\ba}[1]{\begin{array}{#1}}
\newcommand{\ea}{\end{array}}
\begin{document}
\title{Harmonic Superspaces in Low Dimensions}
\author{P.S.Howe and M.I.Leeming\\Department of Mathematics\\King's College,
London}
\date{August 8, 1994}
\maketitle
\begin{abstract}
Harmonic superspaces for spacetimes of dimension $d\leq 3$ are constructed.
Some applications are given.
\end{abstract}
\vfill\eject
%*****************************************************************************
\section{Introduction}
It is well known that the spaces which arise in flat space twistor theory,
including
(complex, compactified) Minkowski space itself, can be constructed as
homogeneous spaces
of the complexified conformal group, which is $SL(4;\com)$ in the four
dimensional case.
In fact, all of the basic twistor spaces are generalized  flag manifolds, which
are, by definition, spaces
of the form $P\backslash G$ where $P$ is a parabolic subgroup of a complex,
simple Lie group, $G$ \cite{ww}.
At the Lie algebra level, a parabolic subalgebra, \gp, of a Lie algebra, \gog,
is one which
contains the Borel subalgebra, \gb, the latter consisting of the Cartan
subalgebra and the
elements of \gog $\ $corresponding to the positive roots. Viewed in this light,
twistor theory can be seen as
a particular application of homogeneous space theory \cite{be}.

This point of view has a natural generalization to supersymmetry with the
complexified superconformal group as a starting point.
The role of flag supermanifolds has been particularly emphasized by Manin
\cite{m}, although
supertwistors were introduced by Ferber \cite{f} and exploited in a field
theory context by Witten \cite{w} and others.
However, in the supersymmetric case, there are a lot more spaces which one can
consider, especially when one has extended supersymmetry.
In fact, two main classes of flag supermanifolds can be constructed: those
whose body
is a conventional twistor space, which we shall refer to as supertwistor
spaces, and those whose body is Minkowski space times an internal
flag manifold which can be thought of as a homogeneous space of the internal
symmetry group
part of the superconformal group. This latter type of superspace has been given
the name harmonic
superspace and was first introduced by GIKOS \cite{o1}\cite{o2} and by Karlhede
{\em et al} \cite{k}. A review of the various superspaces which can be
constructed
in four dimensions is given in \cite{hh}.

In this note we construct all possible harmonic superspaces for spacetimes of
dimension d$\leq$3, which can be obtained as generalized flag supermanifolds of
the superconformal group.
Although there are only a few ordinary twistor spaces in d=3 and none in d=1 or
2
one can still construct a large number of harmonic superspaces. After giving a
brief review
of the general twistor approach, emphasizing the r\^{o}le of double fibrations,
we
construct the various harmonic superspaces which can arise. In the applications
to field theory one is
mainly interested in real, non-compact Minkowski superspace whereas the
corresponding
complex flag supermanifold has a compact body. The non-compact Minkowski
superspace can
be viewed as an open set in this space, and when reality is imposed the
harmonic double fibration is
replaced by a single fibration of harmonic superspace over ordinary Minkowski
superspace.
One then finds that harmonic superspace, although not a complex space, is a CR
supermanifold.
We conclude with a few simple applications.
%*****************************************************************************
\section{General Framework}
Let $X,\ Y$ and $Z$ be complex (super)manifolds and suppose $Y$ fibres over $X$
and over $Z$.
Then we have a double fibration given by the following diagram:

\label{fdf}
\beq
\df{$Y$}{$Z$}{$X$}
\eeq
\vspace{1ex}\\
Information on $X$ can be transferred to $Z$ via the correspondence space $Y$
and vice-versa.
In the applications $X$ will be identified with Minkowski (super)space so
physics can be translated to information on $Z$.
If $x\in X$ then write $\hat{x}=\pi_1\pi_2^{-1}(x)\subset Z$ for the
corresponding subspace in $Z$.
If $z\in Z$ then write $\tilde{z}=\pi_2\pi_1^{-1}(z)\subset X$.
To ensure that a subspace $\hat{x}\subset Z$ corresponds to only one point in
$X$ and that $\tilde{z}\subset X$
corresponds to only one point in $Z$ we demand that $\pi_1$ is 1-1 on the
fibres of $\pi_2$ and $\pi_2$ is 1-1 on the fibres of $\pi_1$.
(This is equivalent to demanding that ($\pi_1,\pi_2):Y\rightarrow(Z,X)$ is an
embedding).

A class of double fibrations can be constructed using homogeneous spaces of a
complex semi-simple Lie Group or a complex, basic, classical, simple Lie
supergroup, $G$.
In fact, if $P$ and $P^\prime$ are parabolic subgroups of $G$ then $P\cap
P^\prime$
is also a parabolic subgroup of $G$ and the following diagram can be
constructed \cite{be}:
\beq
\df{$P\cap P^\prime\backslash G$}{$P\backslash G$}{$P^\prime\backslash G$}
\eeq
\vspace{1ex}\\
$P\cap P^\prime\backslash G$ has typical fibre $P\cap P^\prime\backslash P$
over $P\backslash G$ and typical fibre $P\cap P^\prime\backslash P^\prime$ over
$P^\prime\backslash G$.
We will be interested in $G$ being the superconformal group.
\vspace{2ex}\\
Parabolic subgroups of a complex, semi-simple Lie group, $G$, are constructed
as follows (see e.g. \cite{be}):
\vspace{1ex}\\
Suppose $G$ has Lie algebra \gog. Given a Cartan subalgebra, \gh$\subset$\gog,
and a set of
positive roots, $\Delta^+$(\gog), of \gog, one has a Borel subalgebra,
\gb$\subset$\gog, given by
the direct sum of \gh$\ $and the positive root spaces:
\beq
\r{\gb=\gh}\oplus\bigoplus_{\alpha\in\Delta^+(\r{\gog})}\r{\gog}_\alpha
\eeq
where \gog$_\alpha$ is the root space corresponding to the root $\alpha$.
\\Let \gog$\ $have the simple positive roots
\beq
S=\{\alpha_1,\ldots,\alpha_n\}.
\eeq
Take a subset $S_\r{\gp}\subset S$ and denote by \gl$\ $the set of root spaces
spanned by $S_\r{\gp}$.
Now let \gp$\ $be the direct sum of \gb$\ $and the root spaces corresponding to
the negative roots, $\Delta^-$(\gl), in \gl:
\beq
\r{\gp=\gb}\oplus\bigoplus_{\alpha\in\Delta^-(\r{\gl})}\r{\gog}_\alpha.
\eeq
\gp$\ $contains \gb$\ $therefore it is a parabolic subalgebra of \gog$\ $(by
definition).
Corresponding to \gp$\ $there is a parabolic subgroup $P\subset G$.

The homogeneous space $P\backslash G$ is called a \f{generalized flag
manifold}. Suppose $S\backslash
S_\r{\gp}=\{\alpha_{i_1},\ldots,\alpha_{i_k}\},i_1\leq\cdots\leq i_k$ then
label the
corresponding parabolic subgroup $P_{i_1\cdots i_k}$ and the flag $F_{i_1\cdots
i_k}$.

In the supersymmetric case we will be dealing with a complex Lie supergroup, G,
with basic, classical, simple Lie superalgebra, \gog.
The construction of parabolic subgroups is similar to the construction
described above. The spaces $P\backslash G$ are called \f{generalized flag
supermanifolds}.
Here we have a number of (Weyl) inequivalent Borel subalgebras to choose from.
This leads to a slightly different notation.
If $\alpha_{i_1|j_1},\ldots,\alpha_{i_n|j_n}$ (where $i_l+j_l=l$) are the
simple roots and $S\backslash
S_\r{\gp}=\{\alpha_{i_1|j_1},\ldots,\alpha_{i_k|j_k}\},i_1\leq\cdots\leq
i_k,j_1\leq\cdots\leq j_k$
then label the corresponding parabolic subgroup $P_{{i_1|j_1}\cdots{i_k|j_k}}$
and the flag $F_{{i_1|j_1}\cdots{i_k|j_k}}$.
%*****************************************************************************
\section{Superconformal Groups in Low Dimensions}
The complex conformal groups for d=1,2,3 are $Sp(1),\ Sp(1)\x Sp(1)$ and
$Sp(2)$ respectively.
$Sp(1)$ has only one root, $S=\{\alpha_1\}$. Hence there is only one flag
manifold for d=1, $F_1$:
\beq
F_1=P_1\backslash Sp(1)=\com\proj^1
\eeq
where
\beq
P_1=\left\{\left(\begin{array}{cc} \x & \\\x & \x\end{array}\right)\in
Sp(1)\right\}.
\eeq
\\We identify complex 1d Minkowski space, \mink, via the map
\com$\longrightarrow Sp(1)$ given by
\beq
x\longmapsto s(x):=\left(\begin{array}{cc} 1 & ix\\0 & 1\end{array}\right).
\eeq
If $\bttm a&b\\c&d\ettm\in Sp(1)$ then the right action of $Sp(1)$ on \mink$\
$is easily found to be (where defined, in the non-compact case)
\beq
x\longrightarrow x^\prime=\frac{b+ixd}{a+ixc}.
\eeq
Reality is obtained by demanding that $s^\dagger K_2s=K_2$ where $K_2=\bttm 0
&1\\1 & 0\ettm$;
in this case $x=x^\dagger.$
\vspace{2ex} \\
For d=2 compact, complex Minkowski space is $F_1\x F_1$.
Non-compact complex 2d Minkowski space, \mink, can be identified in $F_1\x F_1$
via the map \com$^2\longrightarrow Sp(1)\times Sp(1)$ given by
\beq
(x^+=x^0+x^1,x^-=x^0-x^1)\longmapsto s(x):=\left(\bttm 1 & ix^+\\0 &
1\ettm,\bttm 1 & ix^-\\0 & 1\ettm\right).
\eeq
The right action of $Sp(1)\times Sp(1)$ on $s(x)$ is the action of the
conformal transformations. Hence $Sp(1)\times Sp(1)$ is the conformal group for
d=2.
Reality is obtained as in the d=1 case.
\vspace{2ex}\\
For d=3, $Sp(2)$ is the relevant group. There are two simple roots,
$S=\{\alpha_1,\alpha_2\}.$
Consider the flag $F_2$:
\beq
F_2=P_2\backslash Sp(2)
\eeq
where
\beq
P_2=\left\{\mb{cccc} \x & \x & & \\\x & \x &  & \\\x & \x & \x & \x\\\x & \x &
\x & \x\me\in Sp(2)\right\}.
\eeq
\vspace{1ex}\\Complex 3d Minkowski space, \mink, can be identified with an open
chart for $F_2$ via the map \com$^3\longrightarrow Sp(2)$ given by
\beq
x\longmapsto s(x):=\bttm 1 & ix^{\alpha\beta}\\0 & 1\ettm
\eeq
where $x^{\alpha\beta}=x^{\beta\alpha},\ \alpha,\beta=1,2.$

The conformal transformations are given by the right action of $Sp(2)$ on
$s(x)$.
Reality is obtained by imposing $s^\dagger K_4s=K_4$ where $K_4=\bttm 0 &
1_2\\1_2 & 0\ettm.$
\vspace{2ex}\\

The superconformal groups are $OSp(N|1),\ OSp(M|1)\x OSp(N|1)$ and $OSp(N|2)$
for d=1,2,3 respectively.
The Lie supergroup $OSp(N|M)$ is given by
\beq
OSp(N\mid M)=\{g\in GL(N\mid M)\mid g^{ST}Jg=J\}
\eeq
where $J=\mb{c|c} S_M & 0\\\hline 0 & K_N\me$ and
\beq
S_M=\bttm 0 & 1_M\\-1_M & 0\ettm,\ K_N=\left\{\begin{array}{ll} {\bttm 0 &
1_n\\1_n & 0\ettm} & \mbox{if $N=2n$}\\\rule[0ex]{0ex}{7ex} {\mb{ccc} 1 & 0 &
0\\0 & 0 & 1_n\\0 & 1_n & 0\me} & \mbox{if $N=2n+1$}\end{array}\right..
\eeq
We will use the system of simple roots for $OSp(N|M)$ given by
\beq
S=\{\alpha_{1|0},\ldots,\alpha_{M|0},\alpha_{M|1},\ldots,\alpha_{M|n}\}
\eeq
For $OSp(N|1)$ we have
$S=\{\alpha_{1|0},\alpha_{1|1},\ldots\ldots,\alpha_{1|n}\}$.
Consider the flag $F_{1|1}$.
\beq
F_{1|1}=P_{1|1}\backslash OSp(N|1)
\eeq
where
\beq
P_{1|1}=\left\{\mb{cc|ccc} \x&&&&\\\x&\x&\x&\ldots&\x\\
\hline \x&&\x&\ldots&\x\\\vdots&&\vdots&&\vdots\\\x&&\x&\ldots&\x\me\in
OSp(N|1)\right\}.
\eeq
1d Minkowski superspace has coordinates $(x,\theta^i),i=1,\ldots,N$ and it can
be identified
in $F_{1|1}$ via the map \com$^{1|N}\longrightarrow OSp(N|1)$ given by
\beq
(x,\theta^i)\longmapsto s(x,\theta):=\mb{cc|c} 1&ix&-\theta^i\\0&1&0\\\hline
0&\theta_i&1_N\me.
\eeq
where the $O(N)$ indices are raised and lowered by $K_N$, i.e.
$\theta^i=K^{ij}_N\theta_j$.

The right action of $OSp(N|1)$ on $s(x,\theta)$ is the action of the
superconformal transformations.
For the real case we impose $s^\dagger J^\prime s=J^\prime$ where
$J^\prime=\mb{c|c} K_2&0\\\hline 0&1_N\me$.
Then $x=x^\dagger$ and $\theta_i=\theta^{i\dagger}$.
\vspace{2ex}\\
For d=2, Minkowski superspace can be identified in
$F(OSp(M|1))_{1|1}\times F(OSp(N|1))_{1|1}$ via the map
\com$^{2|M+N}\longrightarrow OSp(M|1)\times OSp(N|1)$ given by
\beq
(x,\theta)\longmapsto s(x,\theta):=\left(\mb{cc|c}
1&ix^+&-\theta^{+i}\\0&1&0\\\hline 0&\theta^+_i&1_M\me,
\mb{cc|c} 1&ix^-&-\theta^{-i}\\0&1&0\\\hline 0&\theta^-_i&1_N\me\right).
\eeq
In the real case $x^\pm=x^{\pm\dagger},\theta^\pm_i=\theta^{\pm i\dagger}$.
\vspace{2ex}\\
$OSp(N|2)$ has simple roots
$S=\{\alpha_{1|0},\alpha_{2|0},\alpha_{2|1},\ldots,\alpha_{2|N}\}$.
Elements of the parabolic $P_{2|0}$ have the form
\beq
\mb{cccc|ccc}
%% FOLLOWING LINE CANNOT BE BROKEN BEFORE 80 CHAR
%% FOLLOWING LINE CANNOT BE BROKEN BEFORE 80 CHAR
%% FOLLOWING LINE CANNOT BE BROKEN BEFORE 80 CHAR
%% FOLLOWING LINE CANNOT BE BROKEN BEFORE 80 CHAR
%% FOLLOWING LINE CANNOT BE BROKEN BEFORE 80 CHAR
\x&\x&&&&&\\\x&\x&&&&&\\\x&\x&\x&\x&\x&\ldots&\x\\\x&\x&\x&\x&\x&\ldots&\x\\\hline \x&\x&&&\x&\ldots&\x\\\vdots&\vdots&&&\vdots&&\vdots\\\x&\x&&&\x&\ldots&\x\me.
\eeq
Minkowski superspace is identified in $F_{2|0}$ via the map
\com$^{3|2N}\longrightarrow OSp(N|2)$ given by
\beq
(x^{\alpha\beta},\theta^{\alpha i})\longmapsto s(x,\theta):=\mb{cc|c}
1&ix^{\alpha\beta}&-\theta^{\alpha i}\\0&1&0\\\hline 0&\theta^\beta_i&1_N\me.
\eeq
To get reality we demand that $s^\dagger J^{\prime\prime}s=J^{\prime\prime}$
where $J^{\prime\prime}=\mb{c|c} K_4&0\\\hline 0&1_N\me$.
Then $x^{\alpha\beta}=x^{\alpha\beta\dagger}$ and
$\theta^\alpha_i=\theta^{\alpha i\dagger}$.
%*****************************************************************************
\section{Harmonic superspaces}
Here we classify all possible harmonic superspaces for d=1,2,3. They are the
product of
Minkowski superspace with any flag manifold of the internal symmetry group
$O(N)$ (or $O(M)\x O(N)$ for d=2).
\vspace{2ex}\\
\f{d=3:} In the notation of Section 2, if $G=OSp(N|2)$ and $P^\prime=P_{2|0}$
then we have the double fibration
\beq
\df{$P\cap P_{2|0}\backslash OSp(N|2)$}{$P\backslash
OSp(N|2)$}{$P_{2|0}\backslash OSp(N|2)$}
\eeq
\vspace{1ex}\\
We restrict $P_{2|0}\backslash OSp(N|2)$ to the open chart that is identified
with
complex Minkowski superspace, \mink. Now define
\beq
\r{\mink}_H=\pi_2^{-1}(\r{\mink})\subset P\cap P_{2|0}\backslash OSp(N|2),
\eeq
\beq
\r{\mink}_A=\pi_1\pi_2^{-1}(\r{\mink})\subset P\backslash OSp(N|2).
\eeq
\mink$_H$ is called \f{harmonic superspace} and \mink$_A$ is called \f{analytic
superspace}.
Under this restriction the diagram becomes
\beq
\df{\mink$_H$}{\mink$_A$}{\mink}
\eeq
\vspace{1ex}

The essential difference between the supertwistor spaces and the harmonic
superspaces is that
for the former the deviation of the body of $Z$ (in diagram~\ref{fdf}) from the
body of \mink$\ $lies entirely in the spacetime sector
(the $Sp$ part of $OSp$) whereas in the harmonic case it lies entirely in the
internal sector (the $O(N)$ part of $OSp$).

$P$ is chosen so that its even part differs from the even part of $P_{2|0}$
only in the internal sector.
This amounts to choosing a parabolic subgroup of $O(N)$.
$O(N)$ has $n$ simple roots (where $N=2n\ (2n+1)$ if $N$ is even (odd)):
\beq
S=\{\alpha_1,\ldots,\alpha_n\}
\eeq
The flag manifold $F_{i_1\cdots i_k}$ has the real representation
\label{rr}
\beq
F_{i_1\cdots i_k}=\frac{O(N)}{O(N-2i_k)\times U(i_k-i_{k-1})\times\cdots\times
U(i_2-i_1)\times U(i_1)}.
\eeq
(Real representation: $g\in O(N)$ satisfies $g^\dagger g=1$).
\\This is true for $N\neq 4$. When N=4 one finds that the previous equation is
correct for $F_{12}$ and $F_2$ but $F_1=\frac{O(4)}{U(2)}$.

If $P_{i_1\cdots i_k}$ is the chosen parabolic of $O(N)$ then the corresponding
parabolic
of $OSp(N|2)$ is $P=P_{2|i_1\cdots 2|i_k}$, then $P\cap
P^\prime=P_{2|02|i_1\cdots 2|i_k}$.
The choice of parabolic affects the dimension of the odd part of analytic
superspace.
\mink$\ $has $2N$ anticommuting coordinates but \mink$_A$ has only $2(N-i_1)$.
(c.f. the d=4,N=1 case where chiral superspace has half as many coordinates as
ordinary superspace). For example, in $N=3$, there is one possible internal
flag $F_1$ which can be presented as $U(1)\backslash SU(2)$. This case has been
studied in \cite{kz}.

\mink$_H$ has fibre $F_{i_1\cdots i_k}$ over \mink$\ $and as
\mink$_H=\pi_2^{-1}($\mink$),\ $\mink$_H=$\mink$\times F_{i_1\cdots i_k}.$
Hence coordinates for \mink$_H$ are $(x^{\alpha\beta},\theta^{\alpha i},u_I^{\
\ i})$ where $u_I^{\ \ i}$ are homogeneous coordinates for
$F_{i_1\cdots i_k}$ ($I$ is the isotropy, i.e. parabolic, group index).
\mink$_H$ can be identified in $F_{2|02|i_1\cdots 2|i_k}$ via the map
\mink$_H\longrightarrow OSp(N|2)$ given by
\beq
(x^{\alpha\beta},\theta^{\alpha i},u_I^{\ \ i})\longmapsto
s(x,\theta,u):=\mb{cc|c} 1&ix^{\alpha\beta}&-\theta^{\alpha i}\\0&1&0\\\hline
0&\hat{\theta}_I^\alpha&u_I^{\ \ i}\me
\eeq
where $\hat{\theta}_I^\alpha=u_I^{\ \ i}\theta_i^\alpha$.

There is a natural projection from \mink$_H$ to \mink$\ $given by
$(x^{\alpha\beta},\theta^{\alpha i},u_I^{\ \
i})\longmapsto(x^{\alpha\beta},\theta^{\alpha i})$.
The fibre over \mink$_A$ has dimension ${0|2i_1}$. Let coordinates for
\mink$_A$ be
\beq
z=(x_A^{\alpha\beta},\hat{\theta}^{\alpha I},(I\neq 1,\ldots,i_1),v_I^{\ \ i}).
\eeq
The projection from \mink$_H$ to \mink$_A$ is given, in local coordinates, by
\beq
x_A^{\alpha\beta}=x^{\alpha\beta}+i\sum_{I=1}^{i_1}\hat{\theta}^{(\alpha
I}\hat{\theta}^{\beta)}_I,\ \ \hat{\theta}^{\alpha I}=\theta^{\alpha i}u_i^{\ \
I},\ \ (u_i^{\ \ I}=u^{-1}),\ \ v_I^{\ \ i}=u_I^{\ \ i}.
\eeq
The double fibration is
\beq
\dfl{$(x^{\alpha\beta},\theta^{\alpha i},u_I^{\ \
i})$}{$(x_A^{\alpha\beta},\hat{\theta}^{\alpha
I_1+1},\ldots,\hat{\theta}^{\alpha N},u_I^{\ \
i})$}{$(x^{\alpha\beta},\theta^{\alpha i})$}
\eeq

We have covariant derivatives
\\on \mink:
\beq
D_{\alpha\beta}=\partial_{\alpha\beta},\ \ D_{\alpha i}=\partial_{\alpha
i}-i\theta^\beta_i\partial_{\alpha\beta};
\eeq
on \mink$_H$:
\beq
D_{\alpha\beta}=\partial_{\alpha\beta},\ \ \hat{D}_{\alpha I}=u_I^{\ \
i}D_{\alpha i}
\eeq
and the $F_{i_1\cdots i_k}$ coset derivatives;
\vspace{1ex}\\
on \mink$_A$:
\beq
D_{\alpha\beta}=\partial_{\alpha\beta},\ \ \hat{D}_{\alpha I}=u_I^{\ \
i}D_{\alpha i},\ (I=i_1+1,\ldots,n)
\eeq
and the $F_{i_1\cdots i_k}$ coset derivatives.
\vspace{1ex}\\
Note that the coordinates, $z$, on \mink$_A$ regarded as functions on \mink$_H$
satisfy
\beq
\hat{D}_{\alpha I}z=0,\ I=1,\ldots,i_1.
\eeq
Any function, $\phi$, satisfying $\hat{D}_{\alpha I}\phi=0,\ I=1,\ldots,i_1$ is
called analytic.
\vspace{2ex}\\
\f{d=1:} The 1-dimensional case is very similar to the 3d case. The
superconformal group
is $OSp(N|1)$ and the relevant parabolics are $P=P_{1|i_1\cdots 1|i_k}$,
$P^\prime=P_{1|0}$ and $P\cap P^\prime=P_{1|01|i_1\cdots i_k}$.
The double fibration of harmonic superspace over analytic superspace and
Minkowski superspace, written in local coordinates, is
\beq
\dfl{$(x,\theta^i,u_I^{\ \
i})$}{$(x_A,\hat{\theta}^{I_1+1},\ldots,\hat{\theta}^N,u_I^{\ \
i})$}{$(x,\theta^i)$}
\eeq
where
\beq
x_A=x+i\sum_{I=1}^{i_1}\hat{\theta}^I\hat{\theta}_I.
\eeq
\vspace{2ex}\\
\f{d=2:} The 2d case is different in that the coordinates split into 2 sets of
coordinates, each transforming independently
under the superconformal group, $G=OSp(M|1)\times OSp(N|1)$. Hence the complex
superspaces are products of two flags. The double fibration in the compact case
is
\beq
\df{$F_{1|01|i_1\cdots 1|i_k}\times F_{1|01|j_1\cdots 1|j_l}$}{$F_{1|i_1\cdots
1|i_k}\times F_{1|j_1\cdots 1|j_l}$}{$F_{1|0}\times F_{1|0}$}
\eeq
For example, consider the case where $M=N=4,\ k=l=1,\ i_1=j_1=2$, so that
analytic superspace lies in $F_{1|2}\x F_{1|2}$
and harmonic superspace lies in $F_{1|01|2}\x F_{1|01|2}$. It appears that
these superspaces are related to the superfields
recently proposed for $d=2, N=4$ supersymmetry by Lindstr\"{o}m {\em et al}
\cite{l}. For details see \cite{is}.
%*****************************************************************************
\section{Applications of Harmonic Superspace to d=3 Super Yang-Mills}
Suppose we want the fibre of \mink$_H$ over \mink$_A$ to have dimension $0|2k$.
This can be achieved by using a flag $F_{i_1\cdots i_l}$ of $O(N)$ which has
$i_1=k$.
Also it is easier to work with spaces which have as few harmonic coordinates as
possible, so
we require that the flag is minimal, i.e. $F$ has only one subscript.
Hence the relevant flags of $O(N)$ are $F_k,\ 1\leq k\leq n$.

An important theorem in twistor theory, the Ward Observation \cite{ww}, applied
to the double fibration in diagram~\ref{fdf}
which relates holomorphic vector bundles on $Z$ to gauge fields on $X$ can also
be applied in the harmonic case.
It states:
\vspace{2ex}\\
\begin{em}
There is a 1-1 correspondence between holomorphic vector bundles on \mink$_A$
trivial on the fibres of $\pi_2$ and gauge fields on \mink$\ $with vanishing
curvature on $\tilde{z}\subset$\mink$\ \forall\ z\in$\mink$_A$.
\end{em}
\vspace{2ex}

Given a point $z\in$\mink$_A$, translations in the corresponding subset
$\tilde{z}\subset$\mink$\ $
are generated by $\hat{\nabla}_{\alpha I}=u_I^{\ \ i}\nabla_{\alpha i},\
I=1,\ldots,k$ ($\nabla$ is some connection on \mink).
Hence vanishing curvature on all $\tilde{z}\subset$\mink$\ $means
\beq
u_I^{\ \ i}u_J^{\ \ j}\{\nabla_{\alpha i},\nabla_{\beta j}\}=u_I^{\ \ i}u_J^{\
\ j}F_{\alpha i\beta j}=0.
\eeq
This holds for all $u_I^{\ \ i}$. If $k=1$ we have
\beq
u_1^{\ \ (i}u_1^{\ \ j)}F_{\alpha i\beta j}=u_1^{\ \ i}u_1^{\ \ j}F_{\alpha
(i\beta j)}=0
\eeq
and so the symmetric part of the curvature, $F_{\alpha (i\beta j)}$, vanishes.
However, if $k\geq 2$ then the curvature vanishes altogether,
\beq
F_{\alpha i\beta j}=0.
\eeq
(Actually, $u_I^{\ \ i}u_J^{\ \ j}K_{ij}=0$ as $1\leq I,J\leq n$, so $F_{\alpha
i\beta j}=K_{ij}F_{\alpha\beta}$ but $F_{\alpha\beta}$ can be chosen to be 0.
It is just a matter of conventionally choosing the potential $A_{\alpha\beta}$
in terms of $A_{\alpha i}$).

So we have two situations arising. Either $F_{\alpha (i\beta j)}=0$ which is a
curvature constraint that leads to an off-shell Yang-Mills multiplet,
or $F_{\alpha i\beta j}=0$ which is the constraint for Chern-Simons theory.
The Ward observation tells us that holomorphic vector bundles on \mink$_A$
trivial on the fibres of $\pi_2$
correspond to gauge fields on \mink$\ $satisfying either $F_{\alpha (i\beta
j)}=0$ or $F_{\alpha i\beta j}=0$ depending on which harmonic superspace we
have.

So far we have been dealing with complex superspaces. One is really interested
in real Minkowski superspace, M, which
is obtained by imposing $x^{\alpha\beta}=x^{\alpha\beta\dagger},\
\theta^\alpha_i=\theta^{\alpha i\dagger}$ in \mink.
Under this restriction we find that $\pi_1$ becomes 1-1 onto its image and the
double fibration collapses to a single fibration.
It is now convenient to work with the fibres of $\pi_2$ (i.e. the relevant flag
of $O(N)$) in the real representation.We have
\beq
\sif{$M_H=\pi_2^{-1}(M)=M\x \frac{O(N)}{O(N-2k)\x U(k)}$}{$M$}
\eeq
$M_H$ is a CR supermanifold, i.e. a supermanifold with a CR structure.
(The corresponding observation in d=4 was made by Rosly and Schwarz \cite{rs}).
\vspace{2ex}\\
We recall that a \f{CR structure} on a real supermanifold, $N$, of dimension
$2n+m|2n^\prime+m^\prime$ is a complex rank $n|n^\prime$ sub-bundle, $F$, of
the
complexified tangent bundle,$\ T_\r{\bbbold{c}}N=\r{\com}\x N$,
which is involutive (i.e. if $X,Y\in\Gamma(F)$ (sections of $F$) then
$[X,Y]\in\Gamma(F)$).
\vspace{2ex}\\
The Ward observation now becomes:
\vspace{2ex}\\
\begin{em}
There is a 1-1 correspondence between CR vector bundles on $M_H$ (i.e. the
transition functions H satisfy $X(H)=0\ \forall\ X\in\Gamma(\overline{F})$)
trivial on the $\frac{O(N)}{O(N-2k)\x U(k)}$ fibres and gauge fields on $M$
satisfying $F_{\alpha (i\beta j)}=0$ (if $k=1$) or $F_{\alpha i\beta j}=0$ (if
$k>1$).
\end{em}
\vspace{2ex}\\

In the $k=1$ case the resulting off-shell multiplet can be used to construct a
Yang-Mills action for sufficiently low N.
It can also be used to write down an Abelian Chern-Simons action for all $N$.
The $k\geq 2$ case is relevant only for Chern-Simons theory. In some cases we
can go off-shell using the d=4, N=3 Yang-Mills construction of GIKOS \cite{o2}.
For example consider $N=6$. Taking $k=3$ we have the internal flag space
$F_3=\frac{O(6)}{U(3)}$, which is a three dimensional complex space.
By relaxing the constraint on the internal space curvature we find the
off-shell theory, which
involves an infinite number of auxiliary fields due to the dependence on the
harmonic coordinates.
The action is then the Chern-Simons density for the internal curvature
integrated over analytic superspace.
Similarly, the $N=4$ and $5$ cases can be taken off-shell. For these theories
the relevant internal spaces are $\frac{O(4)}{U(2)}$ and $\frac{O(5)}{U(2)}$
respectively.
One encounters a problem in writing down actions when $N>6$. This is because
any parabolic can, at most, reduce the number of
odd coordinates in analytic superspace by $n\ (N=2n$ or $2n+1)$. Hence the
dimension of the measure in the action
is at least $(n-3)$ ($n$ for $d^{2n}\theta$ and $-3$ for $d^3x$). Therefore
when $n>3$ one requires that some of the fields in the integrand have negative
dimension.
\vspace{1ex}\\

Further details and applications will be given elsewhere \cite{hl}.
%*****************************************************************************

\end{document}